\documentclass[conference]{IEEEtran}
\IEEEoverridecommandlockouts
\usepackage{cite}
\usepackage{amsmath,amssymb,amsfonts}
\usepackage{algorithmic}
\usepackage{graphicx}
\usepackage{textcomp}
\usepackage{xcolor}

\usepackage{subcaption}
\def\BibTeX{{\rm B\kern-.05em{\sc i\kern-.025em b}\kern-.08em
    T\kern-.1667em\lower.7ex\hbox{E}\kern-.125emX}}
\begin{document}

\title{A Study on Stock Forecasting Using Deep Learning and Statistical Models}

\author{\IEEEauthorblockN{Himanshu Gupta}
\IEEEauthorblockA{\textit{Tula's Institute} \\
Dehradun, Uttarakhand, India \\
himanshu.code11@gmail.com}
\and
\IEEEauthorblockN{Aditya Jaiswal}
\IEEEauthorblockA{\textit{Graphic Era (Deemed to be University)} \\
Dehradun, Uttarakhand, India
 \\
satendrajaiswal767@gmail.com}
\and

}

\maketitle

\begin{abstract}
Predicting a fast and accurate model for stock price forecasting is been a challenging task and this is an active area of research where it is yet to be found which is the best way to forecast the stock price. Machine learning, deep learning and statistical analysis techniques are used here to get the accurate result so the investors can see the future trend and maximize the return of investment in stock trading. This paper will review many deep learning algorithms for stock price forecasting. We use a record of s\&p 500 index data for training and testing. The survey motive is to check various deep learning and statistical model techniques for stock price forecasting that are Moving Averages, ARIMA which are statistical techniques and LSTM, RNN, CNN, and FULL CNN which are deep learning models. It will discuss various models, including the Auto regression integration moving average model, the Recurrent neural network model, the long short-term model which is the type of RNN used for long dependency for data, the convolutional neural network model, and the full convolutional neural network model, in terms of error calculation or percentage of accuracy that how much it is accurate which measures by the function like Root mean square error, mean absolute error, mean squared error. The model can be used to predict the stock price by checking the low MAE value as lower the MAE value the difference between the predicting and the actual value will be less and this model will predict the price more accurately than other models.
\end{abstract}


\section{Introduction}
Stock Market is a system where currency, stocks, equities, and other financial commodities are either buy and sell by people. Stock market also allows people to own some shares of public company. Traders are mostly wanting to buy that stocks of company which has the potential to increase in the future and refrain that one whose value is expected to fall in the future. Therefore, in this world of financial market it is a demand to forecast the stock price using statistical analysis [1], deep learning, and machine learning technique to maximize the capital gain and decrease the loss. As it is explained that these stock data contain high noise, non-stationary and non-linearity data and to make it stationary we calculate the difference. The noisy data is that which have a gap between a previous stock price and a future price and the non-stationary means that the distribution of the stock price changes during time and non-linearity defines that the feature correlation of stock is various. Investors are one which see the past analysis and there buying and selling depend upon it and it is now also not confirmed that this hypothesis of market is efficient or not. In financial market machine learning and deep learning technique is extensively used for their potential in forecasting market price and predicting the financial market.\\

The stock market is the backbone of any economy and it defines the profit maximization and the minimization in risk. Investing in good stocks can give you good return of interest and as stock prices are linear and that makes it harder to predict and that’s why investors are finding best way to predict the stock prices and in this the machine learning, deep learning and statistical analysis models are used to predict the stock prices. So, in this paper we applied the model of Arima and deep neural network. To get the better result for time series trend we use the Arima model which is the statistical model. The survey also adopts the Recurrent neural network and the type of Recurrent neural network which is Long short-term memory cell to predict the price of the stock data . Our data will also adopt the deep learning algorithm which is convolutional neural network in which we use preprocess and full convolutional neural network [2-3]. Deep learning models provide good result in many areas and detect the dynamics of stock market movement and get good result and in the final evaluation step we considered all the deep learning algorithm and apply it to get the better prediction and then we will see the performance of the model and analyze how it differ with previous study and get to know that how efficient and productive the algorithm is by checking the mean accuracy of all the models.

\section{Literature survey}
This review takes lot of study and analysis into its domain field which is deep learning, the strategy of trading and the stock market data.\\

J.B. Heaton et.al [11] explore deep learning algorithm to solve problem that are arising in financial prediction. It also stated that how deep learning predictor is more efficient and productive in getting result than traditional predictors. It also clarifies that it is easy to handle correlation and how to avoid over fitting.\\

    In this research world it is easily be seen that there are lot of deep learning algorithm.so finding the perfect deep learning algorithm for prediction is a mammoth task. It requires lot of study to find which is the best. Our study also requires to study about different type of algorithm and to see which is the best model. Torkil Aamodt find that Convolutional neural network had efficient result.\\

Mondale et.al[15] study and apply the Arima model to predict the accuracy of stock price prediction. This model was identified by using Akaike information criteria and it is found if there is change in training dataset than the variation in accuracy is little. To determine the accuracy, mean absolute error is the efficient way.
LSTM had been proven successfully for time series prediction. Hengijian jia et al.[13] proposed that the LSTM algorithm learn the stock price pattern in effective way and by applying this approach it gets lower MAE and RMSE value. This study helps to identify this problem as time series prediction and to use sliding window technique to get the better result.\\

Siami-namini[14] in his research found that LSTM overcomes the Arima based model by a large margin and it is approx. 84 and 87 percent. The research also clarifies that when we expand the number of iteration then it does not increase the model efficiency in the stock price prediction.\\

Hiransha[12] used an approach for predicting stock data using RNN, CNN deep learning architecture. This model is capable to detect the trends in stock market. This model gives the better result than the Arima model and in this study, it also clarify that deep learning model is a best and efficient way to predict the time series data.
Literature survey presents that the domain of stock price prediction is yet to be explored at depth as there are many more state of the art techniques that has been proved a better option to predict the stock price.

\section{Methodolgy}
In deep learning it is noted many times that the simplest of algorithm gives the better result than the complex one. While studying about this domain we have keep this in our knowledge and that is the reason we have used multiple deep learning algorithm to predict the time series data.\\

Our main motive is to build an accurate predictive framework. We have used a spy 500 ETF data for building and testing the model. This data has been obtained from the yahoo finance. This dataset includes information about date, open and close value, high and low price of stock and the volume of the stock data. In this survey we have taken the closed price for each stock. Spy data for the period of dec 2015 to Jan 2018 has been used for training and testing of our data. In this we have our data in the form of days so for training record it has 8200 days and for testing it has a range of 8500-10000 days and then we will see which approach gives the least RMSE value for prediction. Using this method, we have built the model where we forecast the closing price of the spy index on weekly basis and monthly basis. That is why for week purpose, we have taken the data from Monday to Friday as these are the only days stock market opens.\\

To make our forecasting more accurate we have used the following models for our study.
\begin{enumerate}
    \item Arima is the statistical analysis model that is mostly used in time series forecasting. In this we must first make our data stationary and then we calculate the significant value of ACF and PACF which is auto correlation functionality and partial auto correlation and then we must set the value for p, d, q value for different stocks.
    \item LSTM is the model which is used in many deep learning models and for time-series forecasting. It is a kind of RNN which capture long term system in time series prediction. Each node in LSTM is connected to previous data streams. It has three gates which is forgot, input and output gate.
    \item In other approaches we have also worked upon deep learning techniques like RNN, CNN which is a neural network and it provides good result in time series prediction. To downside the parameter CNN uses weight sharing and improve efficiency. The input layer is one where data used it for processing and output layer is used for convolutional layer from that it passes through pooling and flattening layer.
\end{enumerate}
These models are used to forecast the stock price and we will see which model is efficient and give the less error.\\

Deep neural network is very popular and making a optimal network in this is difficult task and if we want to see the increment in network performance it is depend upon number of neuron per layer, number of hidden layer, training, activation function, feature set and input data.

\subsection{Data}
As this project is aims to predict stock price or to forecast the time series data. The data that is collected for prediction is a spy 500 data which is an exchange traded fund works under the symbol of spy and we have obtained this stock data from the yahoo finance. It is a day-by-day stock data for all the years of 
Stocks listed on the spy 500 ETF over a period of years [9]. Our data consist of 7000 daily operations and we have removed all the null variables that are there in the data.
\begin{figure}[h!]
    \centering
    \includegraphics[width=0.8\linewidth]{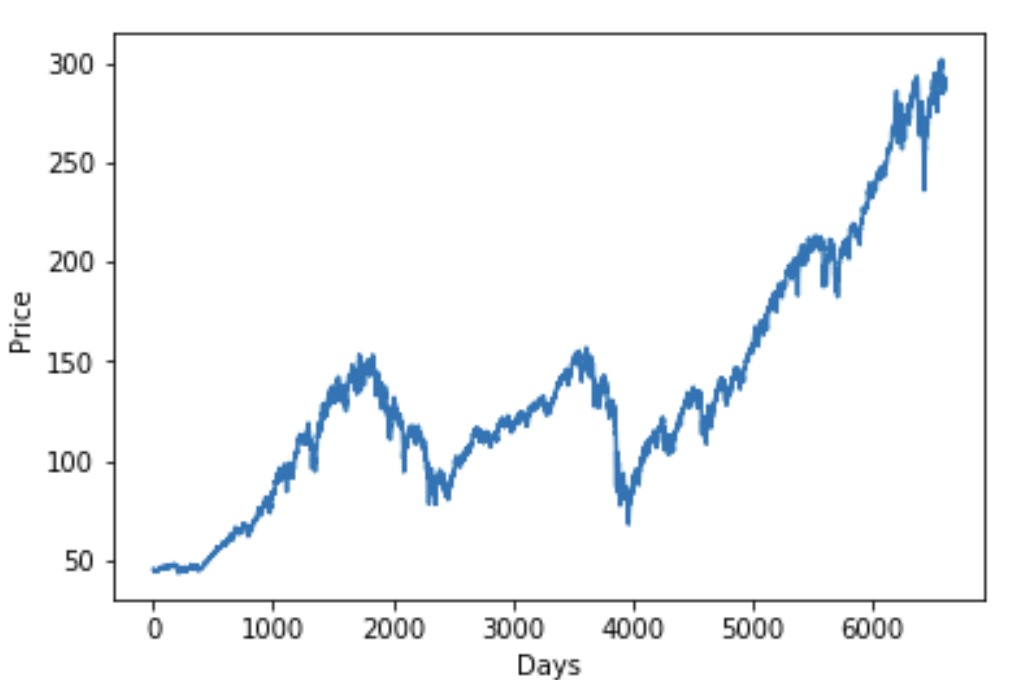}
    \caption{Spy Stock price}
    \label{fig:enter-label}
\end{figure}
\subsection{Model Architecture}
In this research we have use different type of deep learning algorithm to predict the time series data and in the below approaches we will see which of the following gives the efficient result [8].\\

\subsubsection{ARIMA in stock market}
In this approach if we want to predict the time series data than the first condition is that there is the stationary data and it means that the statistical features like mean and other feature like variance does not change over time. Here we use the method which is (p, d, q) and it is used to deal with the non-stationary data. The Augmented Dickey-Fuller and the Null Hypothesis Test is used to check for data non-stationarity so if the NH fails, there is a  probability that the data will be stationary and then we will see if the value of p is less than 0.05 than the NH is rejected and the data is stationary and if the value of p>0.05 than we will use the differencing order this means we will take the value at time (t) and subtract the value at time (t-1) to get the difference [5]. In this term d is the differencing function and for p and q partial autocorrelation and autocorrelation is required and then we will experiment or values by taking the aic value the lower the aic value the better the values of p, d, q are for the model. Now we will plot the residual and check the performance of the model by measuring the mean squared error. \\

\subsubsection{RNN}
In this approach we have used the Recurrent Neural Network deep learning technique A recurrent layer is just a memory cell that computes. Recurrent network defines itself having many cells but it is just one cell used to calculate an output over a set amount of time steps. The cell calculates Y\_0 at X\_0 then moves on to calculating Y\_1 at X\_1 and so on. However, from X\_0 to X\_1 the memory cell produces a state vector. This state factor is used in the next time step as an additional input factor.
\begin{figure}[h!]
    \centering
    \includegraphics[width=0.8\linewidth]{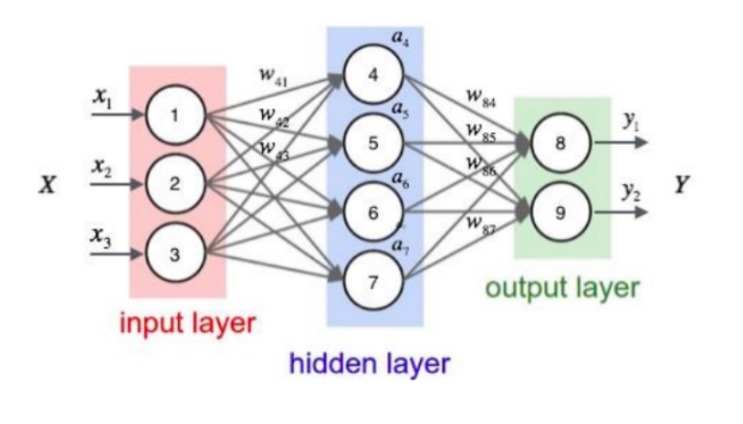}
    \caption{RNN architecture}
    \label{fig:enter-label}
\end{figure}

RNN determines that the current output is also connected to the previous output and hidden layer is connected to it.\\

In this the value s of the hidden layer of RNN not only depends on the current input x, but also to the last hidden layer, Now, form the figure the value of x is for the input layer and the W defines the hidden layer last value and o is the output layer of the model. It uses the error functionality which is back propagation algorithm but it is different as the parameters W, U, and V are the element which got shared in the training process and in the CNN, it is not able to shared and in this approach the output of each step depends on both current and previous network.

\begin{figure}[h!]
    \centering
    \includegraphics[width=0.8\linewidth]{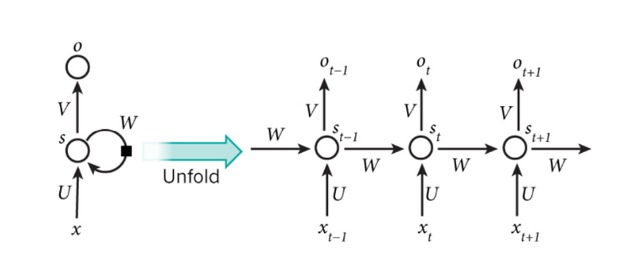}
    \caption{RNN hidden layer}
    \label{fig:enter-label}
\end{figure}

In this instance we are doing sequence to vector which means we ignore all outputs except for the one at the very last time step. This is the default behavior of all recurrent layers in Keras unless return sequences which is equal to True when it is selected. This sequence to vector takes in a batch (128) of time windows and outputs the next time step of window of values. This one at a time output proves to be very slow when training. \\ 

For a faster training convergence, we use a sequence-to-sequence RNN. Compared to the sequence to vector which adjusted the gradient of the loss from the very end of the model all the at layer 2 unit 100, the sequence-to-sequence RNN calculates the loss at each time step and backpropagates the loss from there. Backpropagation is used for training the neural network in which it tunes the weight of the neural network based upon the error rate obtained in the epochs (iteration). This provides much more gradients and speeds up training. It is important to note we still ignore all outputs besides the last one. We just calculate all intermediate values to update the gradient more quickly.\\

\subsubsection{LSTM in stock market}

LSTM is the deep neural network that comes under the family of recurrent neural network. RNN suffers from the problem in which either the network stops its learning or when learn than at a very high learning rate so that it never come close to minimum error. LSTM networks are made in this way that this problem never arises and it becomes suitable for time series data [6-7].  At the beginning of the LSTM model, it has the memory state and it is used for storing information from the past state to the current state. In LSTM there are unit which can add or remove the information from the cell state. Then comes the sigmoid function which is used to forgot the information from the previous state in the cell state. Then, in the input layer we are using a tan-h layer. The output gate, which takes data from previous inputs, it is also considered as predicted value calculated by the model for the current value of slot and it is also decides that which information should be passed on to the hidden state and in the building of the model we have used a optimal learning rate to find the best value for the model to get the accurate result.\\

\subsubsection{CNN in stock market}
CNN is a feed forward neural network and it gives a good performance in artificial intelligence application such as natural language processing, image, and video processing and to the application of time series data [4]. CNN technique which allows the model to improve that how it learns by reducing the parameter value. Input layer is used for processing and the convolutional layer is used for output layer and from that the data passes through pooling layer and flattening layer. Pooling layer is the block of CNN which is used to lower the number of extracted features which reduce the dimensions to speed up the process and if we want to reduce the problem of overfitting than we used the flatten layer which converts the map features into one-dimensional vectors. \\

In preprocess CNN we have a 1D CNN layer that acts as a preprocessor for our SPY data. A 1D-Conv layer acts the same as a 2-D layer used in image processing except that it only operates in one dimension. In image processing the dimensions are height and width, while in our example here we are only worried 1 dimension-- Closing price. A 1D CNN layer does not have any memory as it only calculates an output based on the current filter size. A filter/kernel is just like our moving window. \\

Then we have used a full CNN that apply a wave net architecture. In Wave Net, every layer has kernel size 2 with a stride 1 and using RELU activation functions. However, the 2nd layer uses a dilation rate (used to add parameter into convolutional layer) of 2.\\

This means it skips every other input timestep. The next layer uses a dilation rate of 4 which means it skips every 3 input time steps. The fourth layer uses a dilation rate of 8 which means it skips every 7 timesteps out of 8. This pattern continues which enables the lower layers to learn shorter term patterns while the deeper layers learn the longer-term patterns. The last layer acts as a dense layer to output a single value.

\begin{figure}
    \centering
    \includegraphics[width=1.0\linewidth]{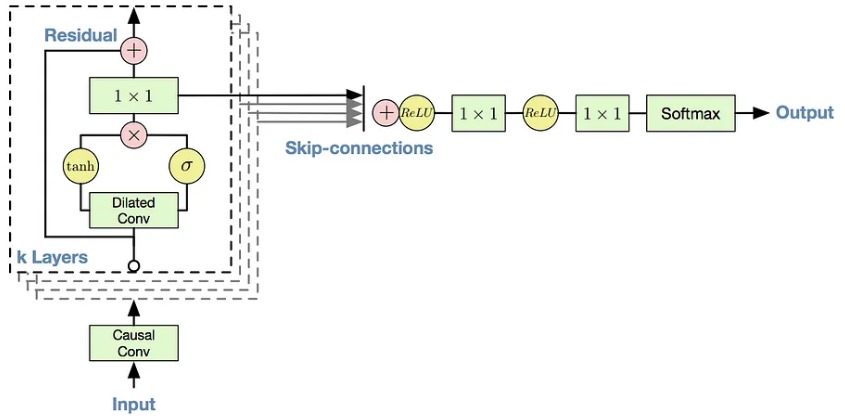}
    \caption{Wave Net Architecture}
    \label{fig:enter-label}
\end{figure}

\section{Experimental results}
\subsection{Experimental setup}
We have used a Jupyter notebook with 8 GB ram or more than it. We have developed our models using python programming language for creating the model. To divide our data into training and testing NumPy is required and for the visualization we have used the matplotlib library.\\

\subsection{Results and Analysis}
In this section we present the results of all the model. A set of features is used here that is a traditional four sets which is open, close, high, low and volume of the stocks. This study includes deep learning models which are ARIMA, RNN, LSTM, CNN-PREPROCESS and FULL CNN and by checking the mean absolute error of all the models we will see that which model gives the effective outcome and productive result as lower the value of mean absolute error than higher will be our deep learning model.\\

In the first model we have used the naïve approach which is to take the previous day data and get the next day result from them and by checking it has given the good MAE value and then we have used the moving average approach which is used to smooth the data and moving average is used to get the desired time span as we have used two models in it which is 20 day moving average and the other one is 5 day moving average and the MAE value of 20 day came higher than the 5 day moving average this means that the price is a recent up trend and we have to play that momentum forward.\\
\\

\begin{figure}[h!]
    \centering
    \includegraphics[width=1\linewidth]{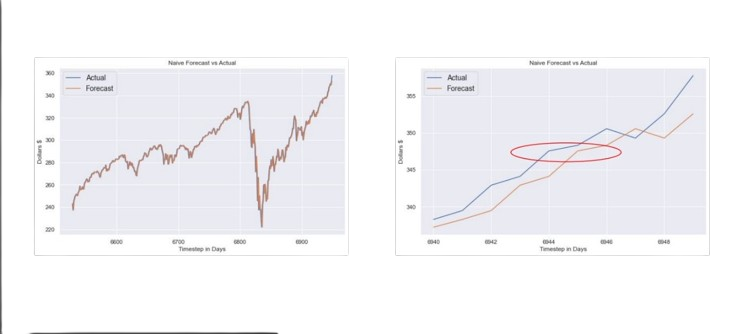}
    \caption{Naïve Forecast}
    \label{fig:enter-label}
\end{figure}

Now we have used the ARIMA model which we have discussed in section 2 about the model architecture and this model gives us the good result and the MAE value of Arima model is 2.8.\\
\\
\\
\\
\\
\\
\\

\begin{figure}[h!]
    \centering
    \includegraphics[width=1\linewidth]{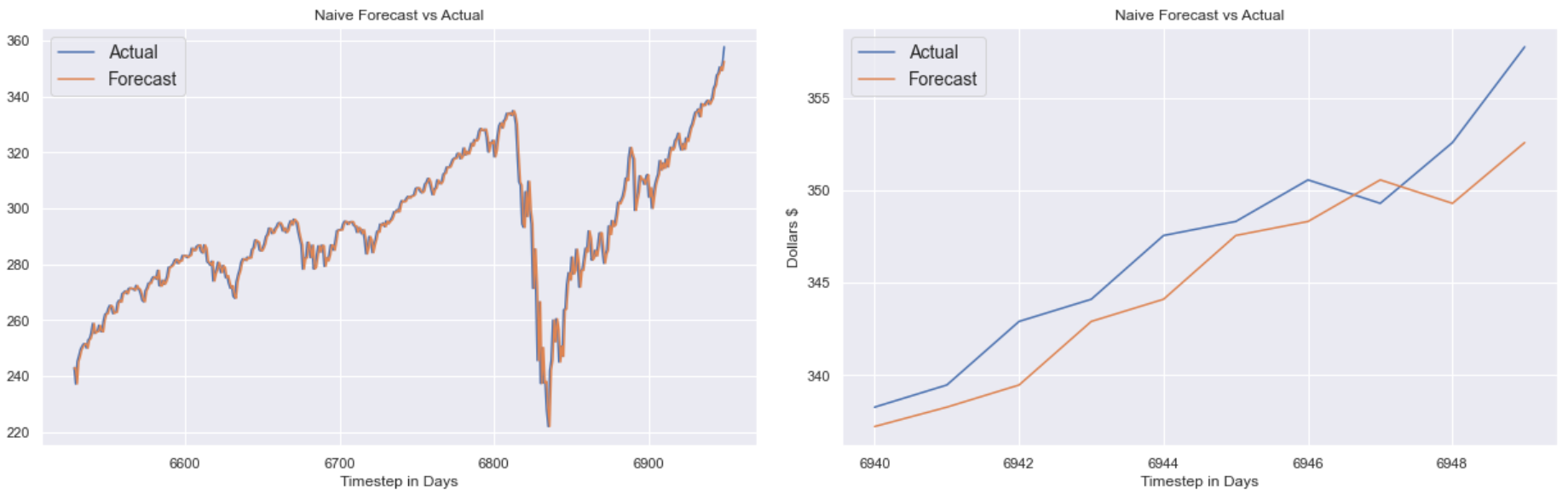}
    \caption{Arima Model}
    \label{fig:enter-label}
\end{figure}

Now we have used a RNN model but with two methods which are sequence to vector and the other one is sequence to sequence and as it has clearly visible from the graph that the seq-to-vector have the MAE value 23.07 which is higher than the MAE value of seq-to-seq which came 4.19. so seq to seq gives the better result and it is efficient to use.
\begin{figure}[h!]
    \centering
    \includegraphics[width=1.0\linewidth]{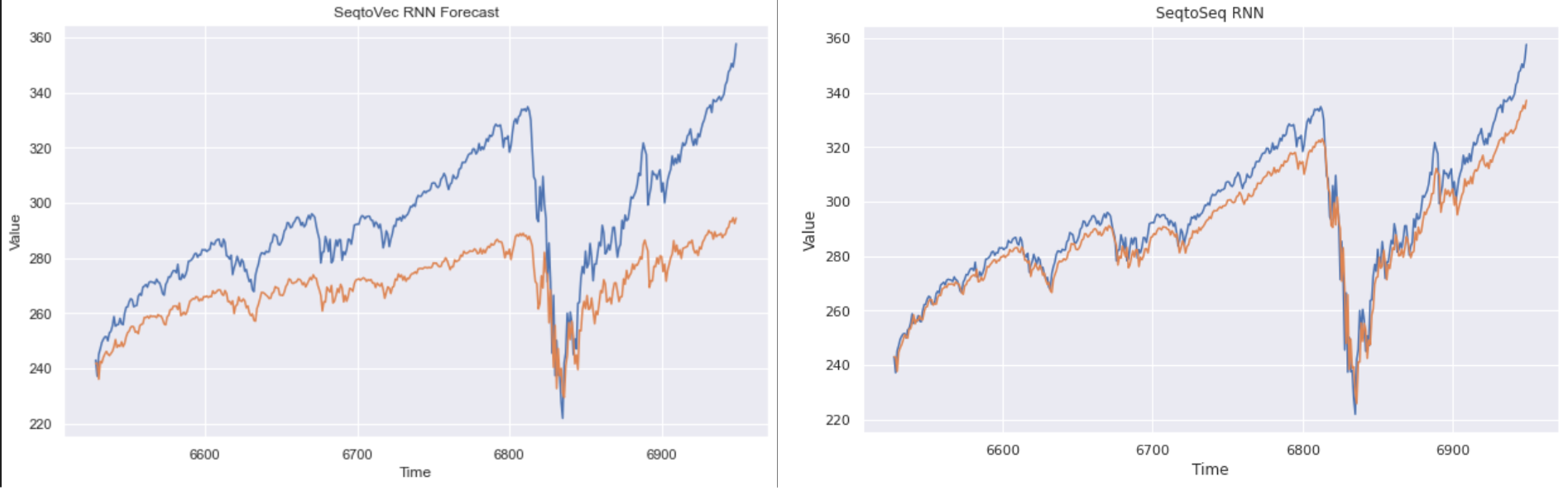}
    \caption{Recurrent Neural Network}
    \label{fig:enter-label}
\end{figure}

The other deep learning model that we have use is LSTM and in this we use two window size one is of 20 and the other one is 30. So, 20-day window size gives the MAE value higher than the 30-day window size. The MAE value of 30-day window size is 1.18 which is lower than any other model which we have used till now.\\
\begin{figure}[h!]
    \centering
    \includegraphics[width=1\linewidth]{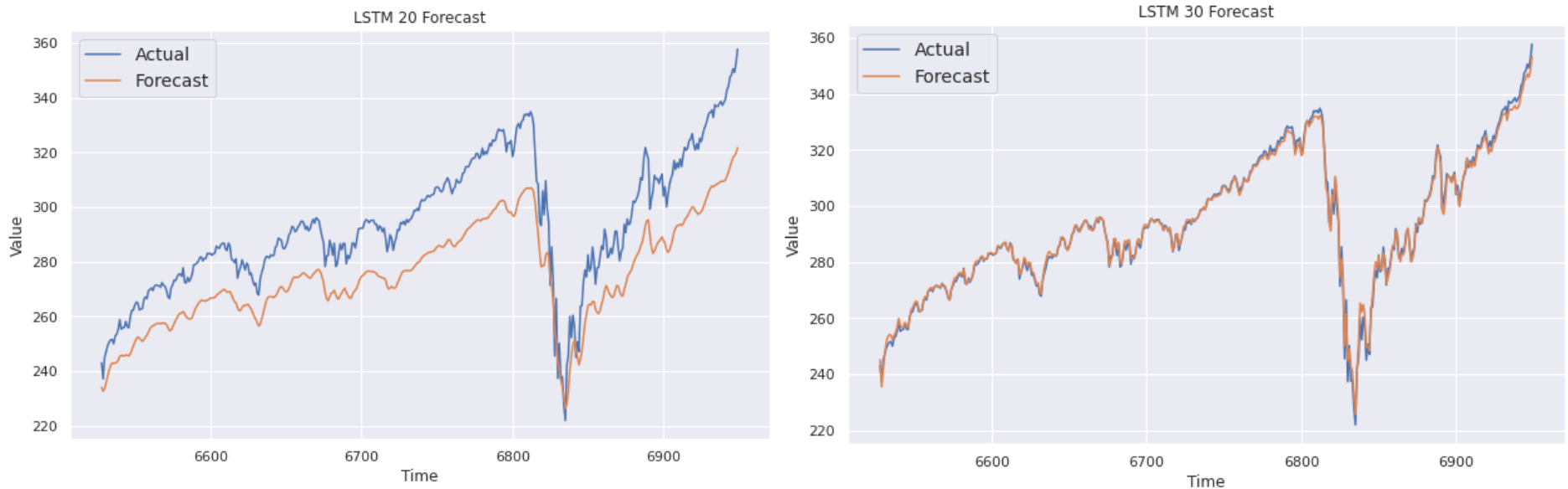}
    \caption{Long Short-Term Memory Model}
    \label{fig:enter-label}
\end{figure}

Now the last deep learning model that we have used is CNN in which there are two methods the first one is pre-process CNN and the other one is Full CNN these methods are discussed clearly in the section 2.4 and the result of preprocess CNN that we got is 12.65 MAE value and for the Full CNN the MAE value is 7.98 which is much lower than the pre process CNN. So, the wave net architecture for full CNN works perfectly and gives the efficient result.\\
\\

\begin{figure}[h!]
    \centering
    \includegraphics[width=1\linewidth]{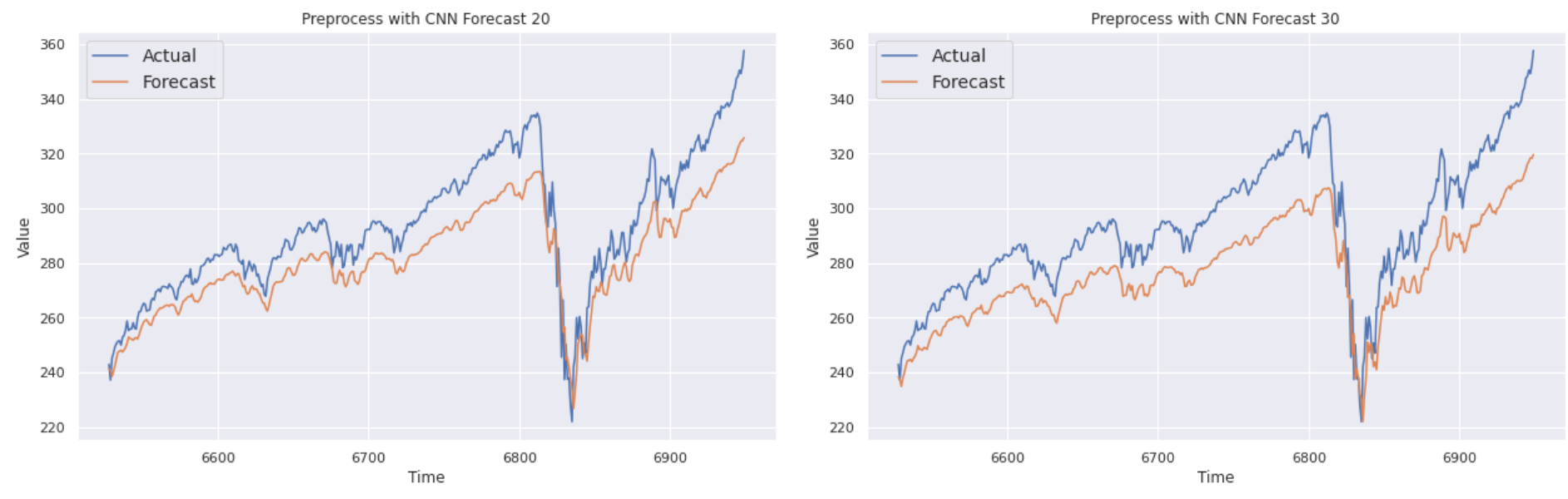}
    \caption{Pre-Process CNN}
    \label{fig:enter-label}
\end{figure}

\begin{figure}[h!]
    \centering
    \includegraphics[width=1\linewidth]{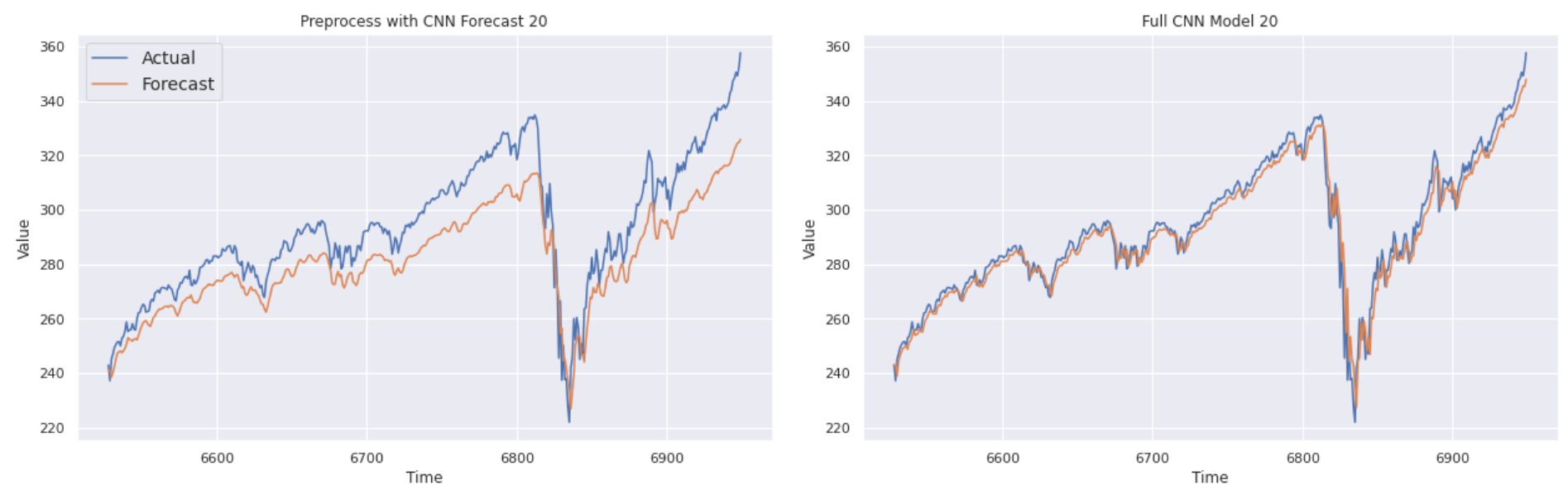}
    \caption{Full CNN}
    \label{fig:enter-label}
\end{figure}

This is the overall result of all the deep learning models that are used in this time series data and predict the stock price prediction and in this survey, we have seen that LSTM model gives us the best result and it is concluded from the above study that this model gives the minimum mean absolute error from all the model and it is the best model which we can use for stock price forecasting.\\
\\

\begin{figure}[h!]
    \centering
    \includegraphics[width=1.0\linewidth]{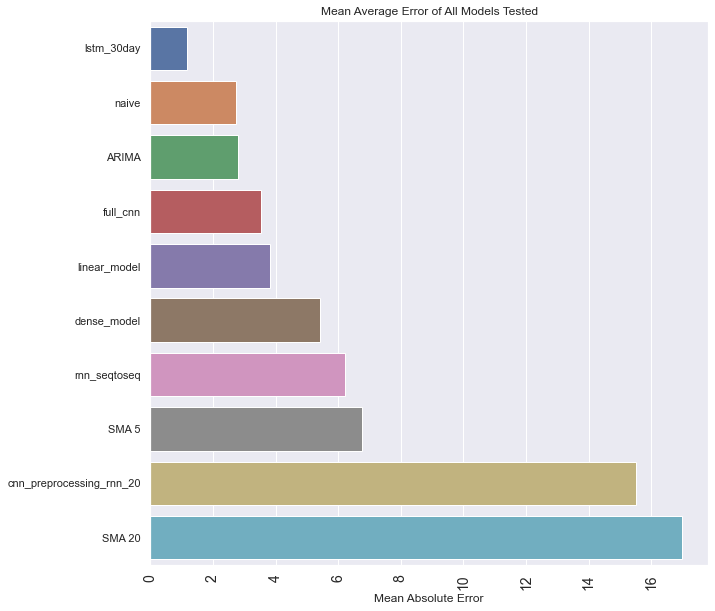}
    \caption{Full Model Results}
    \label{fig:enter-label}
\end{figure}

\section{Conclusion}
Deep Learning and statistical analysis algorithms are nowadays used on modern technology and mainly it is used to form various time series-based forecasting models. Thus, we review the ARIMA, LSTM, RNN, CNN, models and calculate their performance and efficiency. The ARIMA model which is the statistical model predict the result in good way but not as compared to deep learning model and all the other deep learning model does not perform in the same way. The result of all the models is different and the best result is coming by the LSTM model and we have also seen that how the Full CNN outweigh the pre process CNN. This model helps the investors to know when to sell and buy the stock prices. As it is also known that how the stock prices are affected by the climatic or geographical changes, company profit or loss, change in geographical seasons, loss in the region, inflation, competition in the market etc. In the future we will consider these factors also which can change the outcome of stock prices and make our model more accurate.

\section{References}
[1] Forecasting of Indian stock market using time-series ARIMA model
        2014 2nd International Conference on Business and Information Management, ICBIM 2014 (2014), pp. 131-135, 10.1109/ICBIM.2014.6970973\\

[2]  Chen, Zhou and Dai, 2015   K. Chen, Y. Zhou, F. Dai A LSTM-based method for stock returns prediction: A case study of China stock market Proceedings of the IEEE International Conference on Big Data, IEEE Big Data 2015 (2015), pp. 2823-2824, 10.1109/BigData.2015.7364089\\

[3] Chen, Jiang, Zhang and Chen, 2021 W. Chen, M. Jiang, W.G. Zhang, Chen, A novel graph convolutional feature based convolutional neural network for stock trend prediction Information Sciences, 556 (2021), pp. 67-94, 10.1016/j.ins.2020.12.068\\

[4 ] S. Mehtab and J. Sen, “A time series analysis-based stock price prediction using machine learning and deep learning models”, Technical Report, No: NSHM\_KOL\_2020\_SCA\_DS\_1, NSHM Knowledge Campus, Kolkata, INDIA. DOI: 10.13140/RG.2.2.14022.22085/2.\\

[5]  T. Vantuch and I. Zelinka, “Evolutionary based ARIMA models for stock price forecasting”, In Proceedings of Interdisciplinary Symposium on Complex Systems (ISCS’14), pp, 239-247, 2014.\\

[6]   S. Mehtab, J. Sen and A.Dutta, “Stock price prediction using machine learning and LSTM-based deep learning models,” In Proc. of the 2nd Symposium on Machine Learning and Metaheuristic Algorithms and Applications, Chennai, India, October 2020. (Accepted) \\

[7]    X. Shi, Z. Chen, H. Wang, D-Y. Yeung, W-K. Wong, and W-C. Woo, “Convolutional LSTM network: a machine learning approach for precipitation nowcasting,” In Proceedings of the 28th International Conference on Neural Information Processing Systems, vol 1, pp. 802-810, 2015.\\

[8]    J. Sen, “Stock price prediction using machine learning and deep learning frameworks,” In Proceedings of the 6th International Conference on Business Analytics and Intelligence, Bangalore, India, 2018.\\

[9]   Enke, D., Thawornwong, S.: The use of data mining and neural networks for forecasting stock market returns. Expert Syst. Appl. 29(4), 927– 940 (2005).\\

[10]  Hsu, M.-W., et al.: Bridging the divide in financial market forecasting: machine learners vs. financial economists. Expert Syst. Appl. 61, 215– 234 (2016)H., Li, Y., Hu, X., Yang, Y., Meng, Z.,  Chang, K. M. (2013, June). EEG is employed to enhance Massive Open Online Courses Feedback Interaction. In AIED Workshops J. Breckling, Ed., The Analysis of Directional Time Series: Applications to Wind Speed and Direction, ser. Lecture Notes in Statistics.  Berlin, Germany: Springer, 1989, vol. 61.\\

[11] Heaton, J. \& Polson, N. \& Witte, J.. (2016). Deep learning for finance: deep portfolios: J. B. HEATON, N. G. POLSON AND J. H. WITTE. Applied Stochastic Models in Business and Industry. 33. 10.1002/asmb.2209. \\

[12] Hiransha M, Gopalakrishnan E.A., Vijay Krishna Menon, Soman K.P.,NSE Stock Market Prediction Using Deep-Learning Models, Procedia Computer Science, Volume 132,
2018, Pages 1351-1362, ISSN 1877-0509,\\

[13] Jia, Hengjian. (2016). Investigation Into The Effectiveness Of Long Short Term Memory Networks For Stock Price Prediction.\\

[14] Siami Namini, Sima \& Siami Namin, Akbar. (2018). Forecasting Economics and Financial Time Series: ARIMA vs. LSTM. \\

[15] Mondal, Prapanna \& Shit, Labani \& Goswami, Saptarsi. (2014). Study of Effectiveness of Time Series Modeling (Arima) in Forecasting Stock Prices. International Journal of Computer Science, Engineering and Applications. 4. 13-29. 10.5121/ijcsea.2014.4202.

\end{document}